\documentclass[12pt,preprint]{aastex}

\shorttitle{Structure of electrospheres}

\shortauthors{Cheng et al.}

\begin{document}




\title{Structure of the electrospheres  of \\
bare strange stars}







\author{ V.V. Usov\altaffilmark{1}, T. Harko\altaffilmark{2} and K.S. Cheng\altaffilmark{2}}





\altaffiltext{1}{Department of Condensed Matter Physics, Weizmann
Institute, Rehovot 76100, Israel; fnusov@wicc.weizmann.ac.il.}





\altaffiltext{2}{Department of Physics, The University of Hong
Kong, Pokfulam Road, Hong Kong SAR, People's Republic of China;
hrspksc@hkucc.hku.hk, harko@hkucc.hku.hk.}






\begin{abstract}

We consider a thin ($\sim 10^2-10^3$ fm) layer of electrons (the
electrosphere) at the quark surface of a bare strange star, taking
into account the surface effects at the boundary with the vacuum.
The quark surface holds the electron layer by an extremely strong
electric field, generated in the electrosphere to prevent the
electrons from escaping to infinity by counterbalancing the
degeneracy and thermal pressure. Because of the surface tension
and depletion of $s$ quarks a very thin (a few fm) charged layer
of quarks forms at the surface of the star. The formation of this
layer modifies the structure of the electrosphere, by
significantly changing the electric field and the density of the
electrons, in comparison with the case when the surface effects
are ignored. Some consequences of the modification of the
electrosphere structure on the properties of strange stars are
briefly discussed.

\end{abstract}

















\keywords{dense matter --- elementary particles --- stars:
neutron}









\section{INTRODUCTION}

Strange stars made entirely of strange quark matter (SQM) have
long been proposed as an alternative to neutron stars (Alcock,
Farhi, \& Olinto 1986; Haensel, Zdunik, \& Schaeffer 1986;
Glendenning 1996; Cheng, Dai, \& Lu 1998; Weber 1999; Cheng \&
Harko 2000; Alford \& Reddy 2003; Lugones \& Horvath 2003 and
references therein). The possible existence of strange stars is a
direct consequence of the conjecture that SQM, composed of
deconfined $u$, $d$, and $s$ quarks, may be the absolute ground
state of the strong interaction, i.e., it is absolutely stable
with respect to $^{56}$Fe \cite{B71,W84,FJ84}. If SQM is
approximated as a noninteracting gas of quarks, chemical
equilibrium with respect to the weak interaction, together with
the relatively large mass of the $s$ quarks imply that the $s$
quarks are less abundant than the other quarks. Hence in SQM
electrons are required to neutralize the electric charge of the
quarks. The electron density at vanishing pressure is of the order
of $\sim 10^{-4}$ of the quark density \cite{AFO86,KWWG95}.

The electrons, being bounded to SQM by the electromagnetic
interaction alone, are able to move freely across the SQM surface,
but clearly they cannot move to infinity, because of the bulk
electrostatic attraction to the quarks. The distribution of
electrons extends several hundred fermis above the SQM surface,
and an enormous electric field $E\simeq 5\times 10^{17}$
V~cm$^{-1}$ is generated in the surface layer to prevent the
electrons from escaping to infinity, counterbalancing the
degeneracy and thermal pressure \cite{AFO86,KWWG95,HX02}.

The thin layer of electrons with a very strong electric field,
which is called the "electrosphere", drastically affects the
observational appearance of strange stars. First, the
electrosphere of a hot strange star with a bare SQM surface may be
responsible for its thermal emission in a wide range of the
surface temperature $T_S$. The point is that the electric field at
the elecrosphere is a few ten times higher than the critical field
$E_{\rm cr}\simeq 1.3\times 10^{16}$ V~cm$^{-1}$, at which vacuum
is unstable to creation of $e^+e^-$ pairs \cite{S51}. Therefore, a
hot strange star with a bare SQM surface may be a powerful source
of $e^+e^-$ pairs which are created in the electrosphere and flow
away from the star (Usov 1998, 2001a; Aksenov, Milgrom, \& Usov
2003, 2004). Emission of $e^+e^-$ pairs from the electrosphere
dominates in the thermal emission of a bare strange star at
$6\times 10^8~{\rm K}\lesssim T_S\lesssim 5\times 10^{10}$ K,
while below this temperature, $T_S\lesssim 6\times 10^8$ K,
bremsstrahlung radiation of photons from electron-electron
collisions in the electrosphere prevails \cite{JGPP04}. Besides,
the flux of photons generated in the surface layer of SQM via any
mechanism (for example, quark-quark bremsstrahlung) may be
strongly reduced in the process of propagation through the
electrosphere \cite{CH03}. Second, the surface electric field may
be also responsible for existence of a crust of "normal" matter
(ions and electrons) at the SQM surface of a strange star
\cite{AFO86}. This field is directed outward, and the ions in the
inner layer are supported against the gravitational attraction to
the underlying strange star by the electric field.

Recently, it was argued that the properties of the electrospheres
may be changed essentially because of the surface effects
\cite{U04}. It is the purpose of the present paper to study the
structure of the electrospheres by taking into account the surface
effects in details. The different QCD phases of SQM are also
considered. The inclusion of the surface effects drastically
modifies the structure of the electrosphere, by modifying the
surface electric fields and the number density of the electrons.

The remainder of the paper is organized as follows. In \S 2 we
formulate the equations that describe the distributions of the
electric fields and the density of the electrons is the
electrosphere and the boundary conditions. In \S 3 we obtain the
solutions of the equations describing the  properties of the
electrosphere and discuss their properties. Finally, in \S 4 we
summarize our results and discuss some astrophysical applications.

\section{FORMULATION OF THE PROBLEM}

In the electrosphere, electrons are held to the SQM surface by an
extremely strong electric field. The thickness of the
electrosphere is much smaller than the stellar radius $R\simeq
10^6$ cm, and a plane-parallel approximation may be used to study
its structure. In this approximation all values depend only on the
coordinate $z$, where the axis $z$ is perpendicular to the SQM
surface ($z=0$) and directed outward. To find the distributions of
electrons and electric fields in the vicinity of the SQM surface,
we use a simple Thomas-Fermi model considered by Alcock et al.
(1986) and take into account both the finite temperature effects
and the surface tension of SQM as discussed by Kettner et al.
(1995) and Usov (2004), respectively. In the present paper we use
units so that $\hbar =c =k_B =1$. In these units, $e$ is equal to
$\alpha^{1/2}$.

\subsection{Electrostatic equilibrium equations}

The chemical equilibrium of the electrons in the electric field
implies that the value $\mu _\infty =\mu_e -eV$ is constant, where
$V$ is the electrostatic potential, and $\mu_e$ is the electron's
chemical potential. Since far outside the star both $V$ and
$\mu_e$ tend to 0, it follows that $\mu_\infty=0$ and $\mu_e=eV$.

The number density of the electrons is connected with the
electron's chemical potential by the following expression
\cite{KWWG95,CH03}
\begin{equation}
n_e={1\over 3\pi^2}\mu^3_e+ {1\over 3} \mu_eT_S^2= {eV\over
3\pi^2}(e^2V^2 + \pi^2T_S^2)\,,
\label{nezT}
\end{equation}
where $T_S$ is the temperature of the electron layer, which is
assumed constant in the layer and taken equal to the surface
temperature of SQM. This is a very reasonable assumption because
the thickness of the electron layer is very small, and the
electron density is very high. Therefore, the electrons have to be
nearly in thermodynamic equilibrium with the SQM surface.

In the main part of the electrosphere the energy of electrons is
significantly higher than the mass $m_e$, and the
ultrarelativistic approximation when the energy of electrons is
$\sim p$ may be used. In this approximation the Poisson equation
takes the form \cite{AFO86,KWWG95,CH03}:
\begin{equation}
{d^2V\over dz^2}={4\alpha\over 3\pi}[e^2(V^3-V_q^3) +
\pi^2T_S^2(V-V_q)]\,,\,\,z< 0\,, \label{d2V<0}
\end{equation}
\begin{equation}
{d^2V\over dz^2}={4\alpha\over 3\pi}(e^2V^3
+\pi^2T_S^2V)\,,\,\,z>0\,,
\label{d2V>0}
\end{equation}
where $\alpha\simeq 1/137$ is the fine-structure constant, and
$(\alpha/3\pi^2)(e^2V_q^3+\pi^2T_S^2V_q)$ is the quark charge
density.

The boundary conditions for equations (\ref{d2V<0}) and
(\ref{d2V>0}) are $V\rightarrow V_q$ and $dV/dz\rightarrow 0$ as
$z\rightarrow -\infty$, and $V\rightarrow 0$ and $dV/dz\rightarrow
0$ as $z\rightarrow +\infty$, respectively.

To solve the Poisson equation and to find the distributions of
electric fields and electrons it is necessary to know the quark
charge density. In the following we consider the quark charge
density for different QCD phases of SQM with taking into account
the surface effects.

\subsection{Quark charge density in bulk for different QCD phases}

For SQM made of noninteracting quarks the electric charge of the
quarks is positive. Since SQM in bulk has to be electroneutral,
electrons are required to neutralize the electric charge of the
quarks.

The chemical potential ($\tilde\mu_e$) of these electrons at zero
temperature is usually used to characterize the quark charge
density, $\tilde\mu_e=eV_q$. Below, we use only the chemical
potential of electrons for SQM in bulk, and the tilde-sign is
omitted.

It is noted above that in noninteracting SQM the quarks  are
electrically charged because $s$ quarks are more massive than $u$
and $d$ quarks. The mass of $s$ quarks is likely between 50 and
300 MeV. The most traditional estimate is $m_s\simeq 150$ MeV. The
masses of $u$ and $d$ quarks are less than 10~MeV, and we consider
these light quarks as massless particles, $m_u=m_d=0$. Since
$\mu_e\gg m_e$, we neglect the mass of electrons too, $m_e=0$.

Chemical equilibrium under weak reactions imposes
\begin{equation}
\mu_u=\mu-{2\over 3}\mu_e,\,\,\,\,\, \mu_d=\mu_s=\mu +{1\over
3}\mu_e\,, \label{muusd}
\end{equation}
where
\begin{equation}
\mu ={1\over 3}(\mu_u + \mu_d + \mu_s) \label{mumean}
\end{equation}
is the average quark chemical potential.

At low temperatures ($T\ll \mu_e$), in thermodynamic equilibrium
the number density of quarks are
\begin{equation}
N_{u,d}={\mu_{u,d}^3\over \pi^2}\,,\,\,\,\,
N_s={(\mu_s^2-m_s^2)^{3/2}\over \pi^2}\,,\,\,\,\,
N_e={\mu_e^3\over 3\pi^2}\,. \label{Nudse}
\end{equation}

Electrical neutrality requires
\begin{equation}
{2\over 3}N_u -{1\over 3}N_d -{1\over 3}N_s -N_e=0\,. \label{Qe0}
\end{equation}

Equations (\ref{muusd}) - (\ref{Qe0}) yield
\begin{equation}
1-15x+21x^2-44x^3- [(1+x)^2-y^2]^{3/2}=0\,, \label{eqx}
\end{equation}
where $x=\mu_e/3\mu$ and $y=m_s/\mu$. Figure 1 shows a numerical
solution of equation (\ref{eqx}). This solution may be fitted by
the following expression
\begin{equation}
\mu_e=0.248{m_s^2\over \mu}-0.007{m_s^4\over \mu^3}
-0.034{m_s^6\over \mu^5} \label{muefit}
\end{equation}
with the accuracy more than 1\% for $m_s/\mu\leq 1$. The
analytical estimate $\mu_e\simeq m_s^2/4\mu$ performed by Alford
\& Rajagopal (2002) at $m_s/\mu\ll 1$ is well consistent with
equation (\ref{muefit}).

For strange stars, the value of $\mu$ is $\sim 300-350$ MeV at the
surface and $\sim 400-500$ MeV at the center. Taking $m_s\simeq
150$ MeV and $\mu\simeq 300$ MeV as typical parameters of SQM at
vanishing pressure, from equation (\ref{muefit}) we have
$\mu_e\simeq 18.3$ MeV. This value of $\mu_e$ is usually used in
consideration of the properties of noninteracting SQM. However,
the range, where $\mu_e$ may vary, is very wide, from $\sim 2$ MeV
to $\sim 70$ MeV.

It is becoming widely accepted that because of an attractive
interaction between quarks in some specific channels, the ground
state of SQM is a color superconductor (e.g., Bailin \& Love 1984;
Alford, Rajagopal, \& Wilczek 1998; Rapp et al. 1998; Evans et al.
2000; Sch$\ddot{\rm a}$fer 2000; Alford 2001; Alford, Bowers, \&
Rajagopal 2001a). At asymptotic densities ($\gg n_0$), this
superconductor is likely to be in the color-flavor locked (CFL)
phase in which quarks of all three flavors and three colors are
paired in a single condensate, where $n_0\simeq 0.16$ fm$^{-3}$ is
the normal nuclear density. Unfortunately, at intermediate
densities ($\sim 2 n_0$) that are relevant to the SQM surface
layers of strange stars, the QCD phase of SQM is uncertain. In
this low density regime, the SQM may be not only in the CFL phase,
but also in the "two color-flavor superconductor" (2SC) phase in
which only $u$ and $d$ quarks of two color are paired in a single
condensate, while the ones of third color and the s quarks of all
three colors are unpaired. However, it was recently argued that
the density range where the 2SC phase may exist is small, if any
\cite{AR02}.

If rather low temperatures ($T\ll \Delta$) the CFL phase in bulk
is electrically neutral in the absence of any electrons, i.e.,
$\mu_e= eV_q=0$, where $\Delta$ denotes the superconducting gap,
$\Delta \sim 10-10^2$ MeV (Rajagopal \& Wilczek 2001; Steiner,
Reddy, \& Prakash 2002; Weber 2004). The reason for the electrical
neutrality is that BCS-like pairing minimizes the energy if the
quark Fermi momenta are equal. In turn, for equal Fermi momenta,
the numbers of $u$, $d$, $s$ quarks are equal, and the electric
charge of the quarks is zero. This differs qualitatively from the
case of noninteracting SQM. At very high temperatures ($T\sim
\Delta$) the chemical potential of electrons in the CFL phase may
be roughly estimated as $\sim \mu_e \exp\,(-\Delta/T)$, where
$\mu_e$ is given by equation (\ref{muefit}) \cite{W04}.

In the 2SC phase the chemical potential of electrons is, as rule,
higher that the value given by equation (\ref{muefit}) for
noninteractiong SQM, and it may be roughly estimated as
$\mu_e\simeq \mu/4$ (Huang \& Shovkovy 2003 and references
therein). Numerically, for the 2SC phase in bulk at vanishing
pressure we have $\mu_e\simeq 80$ MeV.

In the CFL and 2SC phases, the Cooper pairs are made of quarks
with equal and opposite momenta. Another possibility is a
crystalline color superconductor (CCS), which involves pairing
between quarks whose momenta do not add to zero (Alford, Bowers,
\& Rajagopal 2001b; Bowers \& Rajagopal 2002). Besides, recently
it is shown that the gapless color-flavor locked (gCFL) and
gapless two color-flavor superconductor (g2SC) phases may also
exist in addition to the regular CFL and 2SC phases
\cite{SH03,HS03,AKR04}. In the CCS, gCFL, and g2SC phases,
electrons are present, and it is plausible that their chemical
potential is also in the range from $\sim 10$ MeV to 80 MeV.

\subsection{Thin charged layer at the surface of SQM }

The density of quark states near the surface of SQM is modified
and differs from the density of quark states in bulk
\cite{BJ87,B91, M01}. This results in sharp increase of quark
charge density at the SQM surface. Indeed, the change in number of
quarks of flavor $i$ per unit area because of surface tension is
(Madsen 2000, 2001)
\begin{equation}
n_{i,S}= - {3\over 4\pi}p_{F,i}^2\psi (\lambda_i)\,, \label{niS}
\end{equation}
where
\begin{equation}
\psi (\lambda_i)= \left[{1\over 2}+{\lambda_i\over \pi}-
{1\over\pi}(1+\lambda^2_i) \tan^{-1}(\lambda_i^{-1})\right],
\label{psi}
\end{equation}
$i=\{u,d,s\}$, $p_{F,i}$ is the Fermi momentum of quarks of flavor
$i$, $\lambda_i=m_i/p_{F,i}$, and $m_i$ is the rest mass of quarks
of flavor $i$. The value of $n_{i,S}$ is always negative,
approaching zero for $\lambda_i\rightarrow 0$ (massless quarks).

The rest masses of $u$ and $d$ quarks are very small, and their
densities are not modified significantly by the surface. Thus, the
only appreciable contribution to the surface corrections arises
from the $s$ quarks, i.e., surface effects are highly flavor
dependent. Because of surface depletion of $s$ quarks a thin
charged layer forms at the surface of SQM. The charge per unit
area is positive and equals
\begin{equation}
\sigma = -{1\over 3}en_{s,S}\,. \label{sigma}
\end{equation}

The thickness of the charged layer at the SQM surface of SQM is of
order of $1~{\rm fm}=10^{-13}$ cm, which is a typical strong
interaction length scale.

The thickness of the electron distribution in the electrosphere is
about two order more than the thickness of the charged layer
formed at the surface of SQM because of surface depletion of $s$
quarks (see Alcock et al. 1986; Kettner et al. 1995 and below),
and therefore, we assume that the last is infinitesimal. In this
case, the $z$ component of the electric field
\begin{equation}
E(z)=-dV/dz \label{EzdVz}
\end{equation}
is discontinuous at the SQM surface ($z=0$), and the electric
field jump is
\begin{equation}
\Delta E=E_{\rm ext}(+0)-E_{\rm int}(-0)=4\pi\sigma\,, \label{DE}
\end{equation}
where $E_{\rm ext}(+0)$ and $E_{\rm int}(-0)$ are the
$z-$components of the electric field at the external ($z=+0$) and
internal ($z=-0$) sides of the SQM surface, respectively.

For $m_s\simeq 150$ MeV and $p_{F,s} \simeq 300$ MeV, from
equations (\ref{niS}) - (\ref{DE}) we have $\Delta E\simeq
5.5\times 10^{18}\,{\rm V~cm}^{-1} \simeq 4\times 10^2E_{\rm cr}$
that is about ten times larger than the surface electric field
calculated for SQM in the unpaired phase when the surface effects
are neglected \cite{AFO86,KWWG95,HX02}.

Therefore, the charged layer formed at the surface of SQM because
of surface depletion of $s$ quarks changes the structure of
electrospheres essentially (see below).

The thermal effects have been neglected in the derivation of
equation (\ref{niS}). However, since the thermal energy is small
in comparison with the energy of quarks even if the temperature is
as high as a few ten MeV, this approximation doesn't affect the
conclusions of this paper.

\section{STRUCTURE OF ELECTROSPHERES}

In our study the thermal effects for electrons in the
electrosphere are taken into account because a strange star at the
moment of formation may have the surface temperature $T_S$
comparable with the Fermi energy of electrons inside SQM.

\subsection{Hot electrospheres}

The first integrals of equations (\ref{d2V<0}) and (\ref{d2V>0}),
which satisfy the boundary conditions at $z\rightarrow \pm\infty$,
are $$ {dV\over dz}=\pm\left({2\alpha\over 3\pi}\right)^{1/2}
[e^2(V^4-4V_q^3V+3V_q^4) $$
\begin{equation}
+2\pi^2T_S^2(V-V_q)^2]^{1/2} \label{d1V<0}
\end{equation}
and
\begin{equation}
{dV\over dz}=\pm\left({2\alpha\over 3\pi}\right)^{1/2}
(e^2V^4+2\pi^2T_S^2V^2)^{1/2}\,, \label{d1V>0}
\end{equation}
at $z<0$ and $z>0$, respectively.

Outside of the SQM surface ($z>0$) the $z-$component of the
electric field is directed outward ($E> 0$) to prevent electrons
of the electrosphere from their escaping. Therefore, the sign
minus has to be chosen in equation (\ref{d1V>0}), and the external
electric field is
\begin{equation}
E_{\rm ext}=\left({2\alpha\over 3\pi}\right)^{1/2}
(e^2V^4+2\pi^2T_S^2V^2)^{1/2}\,. \label{Ez>0}
\end{equation}

This equation is valid for any numerical values of $\sigma$ and
$V_q$.

Integration of equations (\ref{EzdVz}) and (\ref{Ez>0}) yields the
electrostatic potential at $z\geq 0$ \cite{CH03}
\begin{equation}
V ={2\sqrt{2}\pi T_S\exp\,[2\sqrt{(\pi\alpha/3)}T_S(z+z_0) \over
e\{\exp\,[4\sqrt{(\pi\alpha /3)}T_S(z+z_0)]-1\}}, \label{VxTS}
\end{equation}
where
\begin{equation}
z_0=\sqrt{3\over \pi\alpha}{1\over 2T_S} \ln \left({\sqrt{2}\pi
T_S\over eV_0}+ \sqrt{1+{2\pi^2T_S^2\over e^2V^2_0}} \right).
\label{z0}
\end{equation}
is a constant of integration, and $V_0$ is the electrostatic
potential at the SQM surface ($z=0$).

Equations (\ref{Ez>0}) - (\ref{z0}) determine the external
electric field as a function of $V_0$ and $T_S$. To find the value
of $V_0$ it is necessary to consider the electric field inside the
SQM surface ($z<0$).

The direction of the internal electric field may be different
depending on the values of $\sigma$ and $V_q$.

At present, the following two cases were considered. In the first
case when $\sigma =0$ and $V_q \neq 0$, the internal electric
field is directed outward similar to the external electric field
\cite{AFO86,KWWG95}. In the second case when $\sigma\neq 0$ and
$V_q=0$, the electric field is symmetric with respect to the SQM
surface [$E(-z)=-E(z)$], i.e., the internal electric field is
directed inward \cite{U04}.

In the last case the electric field is discontinuous at the SQM
surface, and its strength is equal to $\Delta E/2=2\pi \sigma$ at
$z=0$.

Before to consider the internal electric field in a general case
when $\sigma\neq 0$ and $V_q\neq 0$, we find a value of
$\sigma=\sigma_0$ at which the internal electric field is zero for
a given value of $V_q$.

From equations (\ref{EzdVz}) and (\ref{d1V<0}), we have a solution
$E_{\rm int}=0$ if $V=V_q$ at $z < 0$. At the SQM surface the
electrostatic potential $V$ is continuous, and therefore,
$V_0=V_q$ for this solution, while the electric field is
discontinuous, $\Delta E=E_{\rm ext}(+0)= 4\pi\sigma_0$.
Substituting $E_{\rm ext}(+0)$ from this equation into equation
(\ref{Ez>0}) where $V=V_q$ we get
\begin{equation}
\sigma_0={1\over 4\pi}\left({2\alpha\over 3\pi}\right)^{1/2}
(e^2V_q^4+2\pi^2T_S^2V_q^2)^{1/2}\,. \label{sigma0}
\end{equation}

To characterize importance of the surface effects for generation
of strong electric fields in the vicinity of the SQM surface we
introduce the following parameter:
\begin{equation}
\eta ={\sigma\over\sigma_0}={\Delta E \over E_q}\,, \label{eta1}
\end{equation}
where $E_q$ is the strength of the external electric field given
by equation (\ref{Ez>0}) at $V=V_q$. The surface effects are
unessential for the elecrosphere structure if the jump of the
electric field $\Delta E$ is much smaller than the electric field
near the SQM surface, i.e., if $\eta\ll 1$. At $\eta \geq 1$ the
electric field structure in the vicinity of the SQM surface
changes qualitatively in comparison with the field structure
considered by Alcock et al. (1986) and Kettner et al. (1995) when
the surface effects have been ignored ($\eta =0$). For example, at
$\eta > 1$ the internal electric field is directed inward ($E_{\rm
int}<0$), while at $\eta =0$ it is directed outward ($E_{\rm
int}>0$). Besides, in the former case the electric field and the
density of electrons in the electrosphere increase significantly
(see below).

Equations (\ref{niS}), (\ref{sigma}), (\ref{sigma0}) and
(\ref{eta1}) yield
\begin{equation}
\eta= \eta_0 \left(1+{2\pi^2 T_S^2\over e^2V_q^2}\right)^{-1/2},
\label{eta2}
\end{equation}
where
\begin{equation}
\eta_0= \left({3\pi\alpha\over 2}\right)^{1/2} \left({p_{F,s}\over
eV_q}\right)^2\psi(\lambda_s)\,. \label{eta0}
\end{equation}

For the most conservative parameters, $m_s\simeq 150$ MeV,
$p_{F,s}\simeq 300$ MeV and $eV_q\simeq 20$ MeV, from equations
(\ref{eta2}) and (\ref{eta0}) we have $\eta_0\simeq 9$ and
\begin{equation}
\eta\simeq 9\left[1+\left(T_S\over 4.5~{\rm MeV}
\right)^2\right]^{-1/2}. \label{eta3}
\end{equation}

From this equation we can see that if the surface temperature is
not extremely high ($T_S< 40$ MeV), $\eta$ is $>1$, i.e., the
surface depletion of $s$ quarks is responsible mainly for the
generation of extremely strong electric fields near the SQM
surface, not the electrons of SQM in bulk.

Using equations (\ref{muusd}), (\ref{muefit}), (\ref{psi}),
(\ref{eta2}) and (\ref{eta0}), we performed numerical calculations
of $\eta$ for unpaired SQM at different temperatures, $0\leq
T_S\leq 40$ MeV. Figure 2 shows $\eta$ as a function of $m_s/\mu$
for $\mu =300$ MeV and $50\leq m_s \leq 300$ MeV. From Figure 2 we
can see that $\eta$ is in the range from $\sim 3\times 10^2$ at
$m_s\simeq 50$ MeV and low temperatures ($T_S\lesssim {\rm a~few}$
MeV) to $\sim 0.1$ at $m_s\simeq 300$ MeV and $T_S\simeq 40$ MeV.

Hence, both the cases ($\eta > 1$ and $\eta < 1$) may occur in the
electrospheres of bare strange stars depending on the surface
temperature and the SQM parameters. Since the internal electric
field is directed outward at $\eta <1$ and inward at $\eta >1$,
the sign $-$ or $+$ has to be taken in equation (\ref{d1V<0}) at
$\eta <1$ or $\eta >1$, respectively.

The sign of $1-\eta$ coincides with the sign of $V-V_q$ at $z\leq
0$, and from equations (\ref{EzdVz}) and (\ref{d1V<0}) the
internal electric field may be written in the following form:
$$E_{\rm int}= -{dV\over dz}=\left({2\alpha\over
3\pi}\right)^{1/2}(V_q-V)$$
\begin{equation}
\times [e^2(V+V_q)^2 + 2e^2V_q^2+2\pi^2T_S^2]^{1/2} \label{dVdzf}
\end{equation}
that is valid for any values of $\eta$.

Integration of equation (\ref{dVdzf}) yields the electrostatic
potential at $z\leq 0$:
\begin{equation}
V=\left[{{(1+\xi\coth\chi )^2-\xi^2 +1 \over 1+\xi\coth\chi}-1}
\right]V_q\,, \label{VzT<0}
\end{equation}
where
\begin{equation}
\xi=\left[{3\over 2}\left(1+{\pi^2T_S^2\over 3e^2V_q^2}\right)
\right]^{1/2}\,, \label{xi}
\end{equation}
\begin{equation}
\chi =\exp\left[\left({2\alpha\over 3\pi}\right)^{1/2} eV_q\xi
(|z|+\tilde z_0)\right]\,, \label{chi}
\end{equation}
$$\tilde z_0=\left({3\pi\over 2\alpha}\right)^{1/2}
{1\over \xi eV_q} \,\ln \,\coth ^{-1}$$
\begin{equation}
\left[ {1\over 2\xi}\left({V_0\over V_q}-1\right)+ {1\over
2\xi}\sqrt{\left({V_0\over V_q}+1 \right)^2+4(\xi^2-1)}\right]
\label{z0int}
\end{equation}

Substituting $\sigma$, $E_{\rm ext}(+0)$, and $E_{\rm int}(-0)$
from equations (\ref{sigma}), (\ref{Ez>0}), and (\ref{dVdzf}) into
equation (\ref{DE}), we have the following equation for the
electrostatic potential $V_0$ at $z=0$:
$$(V_0-V_q)[e^2(V_0+V_q)^2 + 2e^2V_q^2+2\pi^2T_S^2]^{1/2}$$
\begin{equation}
+(e^2V_0^4+2\pi^2T_S^2V_0^2)^{1/2}= (3\pi/2)^{1/2}p^2_{F,s}\psi
(\lambda_s )\,. \label{V0SQM}
\end{equation}

Figure~3 shows the value of $V_0$ at the surface of unpaired SQM
as a function of $m_s/\mu$ for $\mu\simeq 300$ MeV and different
values of $T_S$.

Equations (\ref{nezT}), (\ref{Ez>0}) - (\ref{z0}) and
(\ref{dVdzf}) - (\ref{V0SQM}) determine the electric field and the
density of electrons in the electrosphere of a bare strange star
as functions of $m_s$, $\mu$, $V_q$ and $T_S$. In turn, $V_q$ is
not a free parameter and may be calculated for different QCD
phases of SQM as functions of $m_s$, $\mu$ and $T_S$ (see \S 2).
From Figure~4 we see that the external field at the SQM surface is
extremely high [$\sim (1-7)\times 10^2\,E_{\rm cr}$] that is about
ten times more than the same without taking into account the
surface effects. Figures 5 and 6 show the profiles of the
electrostatic potential of electrons and the electron density in
the vicinity of the SQM surface, respectively.

\subsection{Cold electrospheres}

In $\sim 10$ s after formation of a bare strange star the surface
temperature is \cite{PU02}
\begin{equation}
T_S\ll T_* = {1\over \pi}\, eV_0\sim 10\,{\rm MeV}\,.
\label{TSlimit}
\end{equation}

From equations (\ref{nezT}) - (\ref{d2V>0}) it follows that in
this case the thermal effects are not essential for the
distributions of electrons and electric fields near the SQM
surface. Even if the surface of a bare strange star is heated
occasionally by accretion of matter, for example, so that the
thermal luminosity of the star is as high as $\sim
10^{44}-10^{45}$ ergs~s$^{-1}$, the surface temperature is
$T_S\simeq (1-2)\times 10^9~{\rm K}\ll T_*$ \cite{U01b}. Hence,
strange stars in the process of their evolution are mostly at the
state when $T_S\ll T_*$, and therefore, we discuss this low
temperature case especially.

In this case the electrostatic potential at $z\geq 0$ has a very
simple form:
\begin{equation}
V=\left({3\pi\over 2\alpha}\right)^{1/2}{1\over e(z+z_0)} ={C\over
e(z+z_0)}\,, \label{VTzeroz>0}
\end{equation}
where
\begin{equation}
z_0=\left({3\pi\over 2\alpha}\right)^{1/2}{1\over eV_0} ={C\over
eV_0} \label{z0Tzeroz>0}
\end{equation}
is the typical thickness of the electrosphere, and $C=(3\pi
/2\alpha)^{1/2}=5.013\times 10^3$ MeV~fm. For $eV_0=40$ MeV, we
have $z_0\simeq 125$ fm.

From equations (\ref{nezT}), (\ref{Ez>0}) and (\ref{VTzeroz>0})
the external electric field and the density of electrons are,
respectively,
\begin{equation}
E_{\rm ext}={C\over e(z+z_0)^2}\,, \label{EextTzeroz>0}
\end{equation}
\begin{equation}
n_e={1\over 3\pi^2}\left({3\pi\over 2\alpha}\right)^{3/2} {1\over
(z+z_0)^3}\,. \label{neTzeroz>0}
\end{equation}

The electrostatic potential inside the SQM surface is not
simplified significantly at low temperatures, and it is given by
equations (\ref{VzT<0}), (\ref{chi}) and (\ref{z0int}) where
$\xi=(3/2)^{1/2}$.

From equations (\ref{eta0}) and (\ref{V0SQM}), in the limit of
zero temperature the equation for $V_0$ is
\begin{equation}
(V_0-V_q)[(V_0+V_q)^2+2V_q^2]^{1/2} + V_0^2=\eta_0 V_q^2,.
\label{V0T0}
\end{equation}

This equation has the following solution \begin{equation}
V_0=\left[{1\over \eta_0}\pm\left({1\over\eta_0^2}
+{\eta_0^2-3\over 2\eta_0}\right)^{1/2}\right]V_q\,, \label{V0f}
\end{equation}
where the sign $+$ or $-$ has to be taken at $\eta \geq 1$ or
$\eta <  0$, respectively.

For $\eta_0\simeq 9$, which relates to the most conservative
parameters of SQM (see above), from equation (\ref{V0f}) we have
$V_0\simeq 2.2V_q$ that is 2.93 times more than $V_0=(3/4)V_q$
found in many papers (e.g., Alcock et al. 1986; Kettner et al.
1995) where the surface effects were ignored. Then, from equations
(\ref{nezT}) and (\ref{Ez>0}) taken at $T_S=0$ we find that at the
SQM surface the external electric field ($E_{\rm ext}\propto
V_0^2$) and the density of electrons ($n_e\propto V_0^3$) increase
because of the surface effects by factors 8.6 and 25,
respectively, and they are $E_{\rm ext}(z=0)\simeq 4\times
10^{18}~{\rm V~cm}^{-1}\simeq 3 \times 10^2 E_{\rm cr}$ and
$n_e(z=0)\simeq 3.7\times 10^{35}$ cm$^{-3}$.

\section{DISCUSSION}

In this paper, we have considered the structure of extremely
strong electric fields and the distribution of the density of
electrons in the vicinity of the SQM surface of a bare strange
star.

 It is taken into account that a very thin (a few fm) charged
layer forms at the SQM surface because of surface depletion of $s$
quarks. We have shown that the surface charged layer changes
significantly both the field structure and the density of
electrons as compared with the same calculated in many papers
(e.g., Alcock et al. 1986; Kettner et al. 1995; Hu \& Xu 2002;
Cheng \& Harko 2003), where the surface effects are ignored. These
changes are especially large if the surface temperature is not
very high ($T_S\lesssim$ a~few MeV).

For a bare strange star the structure of electric fields and the
density of electrons near the stellar surface depend on $m_s$,
$\mu$, and $T_S$ [see equations (\ref{nezT}), (\ref{Ez>0}) -
(\ref{z0}), (\ref{dVdzf}) - (\ref{V0SQM}) and \S2]. For SQM at
vanishing pressure the value of $\mu$ is about 300 MeV with
accuracy of about 20\%, and the main uncertainty of the parameters
of the electrospheres of bare strange stars is because of
uncertainty of $m_s$ that is estimated as 150 MeV within a factor
of 2-3 or so.

The density of electrons in the electrosphere of a bare strange
star increases because of the surface effects by a factor of few
ten (see \S~3). This may change significantly the thermal emission
from the stellar surface. In the Introduction, the main mechanisms
of the thermal emission are mentioned. They are creation of
$e^+e^-$ pairs in a supercritical electric field \cite{U98},
quark-quark bremsstrahlung from the surface layer of SQM
\cite{CH03}, and electron-electron bremsstrahlung from the
electron layer \cite{JGPP04}. In supercritical electric fields,
$E\gg E_{\rm cr}$, the rate of pair production when electrons are
created into the empty quantum states is extremely high
\cite{S51}, and all the empty states are occupied by creating
electrons practically instantly. Then, the rate of pair production
in the electrosphere is determined by the process of
thermalization of electrons which favors the empty-state
production \cite{U98}. Using this, we have roughly the following
dependence of the strange star luminosity in $e^+e^-$ pairs on
$V_0$:
\begin{equation}
L_\pm\propto\cases{{\rm constant} & at $T_S < 0.01\,eV_0$,\cr
V_0^{1.8}& at $0.01\,eV_0\lesssim T_S\lesssim 0.1\,eV_0$,\cr V_0^3
& at $T_S > 0.1\,eV_0$\,,\cr} \label{Lpm}
\end{equation}
where we use $V_0$ which is connected with the density of
electrons at the electrosphere [see equation (\ref{nezT})].

For the most conservative parameters of SQM ($m_s\simeq 150$ MeV
and $\mu\simeq 300$ MeV), the value $V_0$ increase $\sim 3$ times
because of the surface effects. Hence, at high temperatures
$L_\pm$ may increase up to a few ten times in comparison with the
value calculated by Usov (1998) for the electrosphere model where
the surface effects are ignored.

It has been shown by Cheng \& Harko (2003) that quark-quark
bremsstrahlung (and another radiation from SQM) may be strongly
suppressed in the process of its propagation through the
electrosphere. Most probably, the increase of the density of
electrons by a factor of $\sim 20-30$ because of the surface
effects results in almost complete suppression of the outgoing
radiation from SQM if the surface temperature is not very high
($T_S < 1~{\rm MeV}$). In this case practically all radiation of a
bare strange star is generated by the electron layer at $z>0$. In
contrary, we expect that the electron density increase doesn't
affect significantly the electron-electron bremsstrahlung
radiation from the electrosphere at rather low temperature ($T_S <
0.1~{\rm MeV}$) where this radiation prevails. Indeed, in this
case the bremsstrahlung radiation is generated mainly at the
distance from the SQM surface where the local plasma frequency of
electrons is equal to the frequency of radiation (Jaikumar et al.
2004). Since at any temperature the plasma frequency of electrons
is connected directly with the density of electrons, the density
of electrons in the radiating region is more or less the same
irrespective of the structure of the electron layer. Therefore,
the electron-electron bremsstrahlung radiation at a given
frequency is more or less the same, too. We hope to deal with
detail calculations of the thermal emission from a bare strange
star for the electrosphere model developed in this paper
elsewhere.

Recently, the cooling of young bare strange stars has been studied
numerically using the electrosphere model where the surface
depletion of $s$ quarks is ignored \cite{PU02}. It was shown that
the thermal luminosity of such a star in photons and $e^+e^-$
pairs may be up to $\sim 10^{49}- 10^{50}$ ergs~s$^{-1}$ for a few
second after the star formation and remains high enough ($\gtrsim
10^{36}$ ergs~s$^{-1}$) as long as the surface temperature is
higher than $\sim 0.1$ MeV. The increase of the density of
electrons in the electrosphere because of the surface effects has
to modify the light curves calculated by Page \& Usov (2002). We
expect that at the first stage of the strange star cooling, when
the neutrino losses dominate, the thermal radiation from the
stellar surface increases in accordance with equation (\ref{Lpm}),
i.e., by a factor of few ten at $T_S > 0.1\,eV_0$. At the second
stage, when the losses in the surface thermal radiation prevails,
the thermal luminosity decreases more fast than in the case of the
non-modified electrosphere, especially at low temperatures ($T_S <
0.1$ MeV) where the main mechanism of the thermal emission is the
electron-electron bremsstrahlung that is ignored by Page \& Usov
(2002).

At the surface of a strange star a massive normal matter crust may
be formed by accretion of matter onto the star. From Figure~3 we
can see that at rather low temperatures ($T_S\ll T_*\sim 10$ MeV)
the electrostatic potential of electrons $eV_0$ is more than the
electron chemical potential ($\sim 25$ MeV) at which neutron drip
occurs \cite{BPS71}. Therefore, the maximum density of the crust
is limited by neutron drip and is about $4.3\times 10^{11}$
g~cm$^{-3}$ \cite{AFO86}. In this case, the maximum mass of the
crust is $\sim 10^{-5}M_\odot$.

\begin{acknowledgements}

This work was supported by a RGC grant of the Hong Kong
Government. V.V.U. would like to thank the Department of Physics,
University of Hong Kong, where this work was carried out in part,
for its kind hospitality. The research of V.V.U. was supported by
the Israel Science Foundation of the Israel Academy of Sciences
and Humanities.

\end{acknowledgements}

\onecolumn

\vspace{0.2in}

\begin{figure}[h]
\plotone{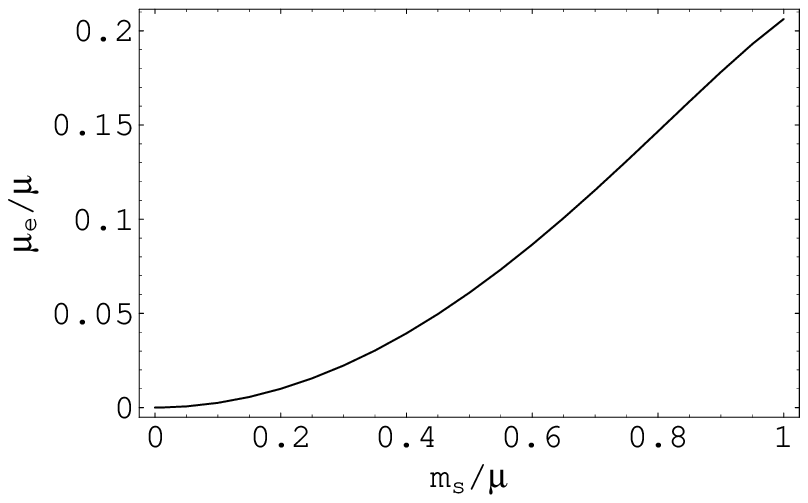} \caption{Electron chemical potential $\mu_e$ for
unpaired SQM in bulk as a function of the $s$ quark mass $m_s$.
Both these values are measured in the average quark chemical
potential $\mu$.} \label{FIG1}
\end{figure}

\vspace{0.2in}

\begin{figure}[h]
\plotone{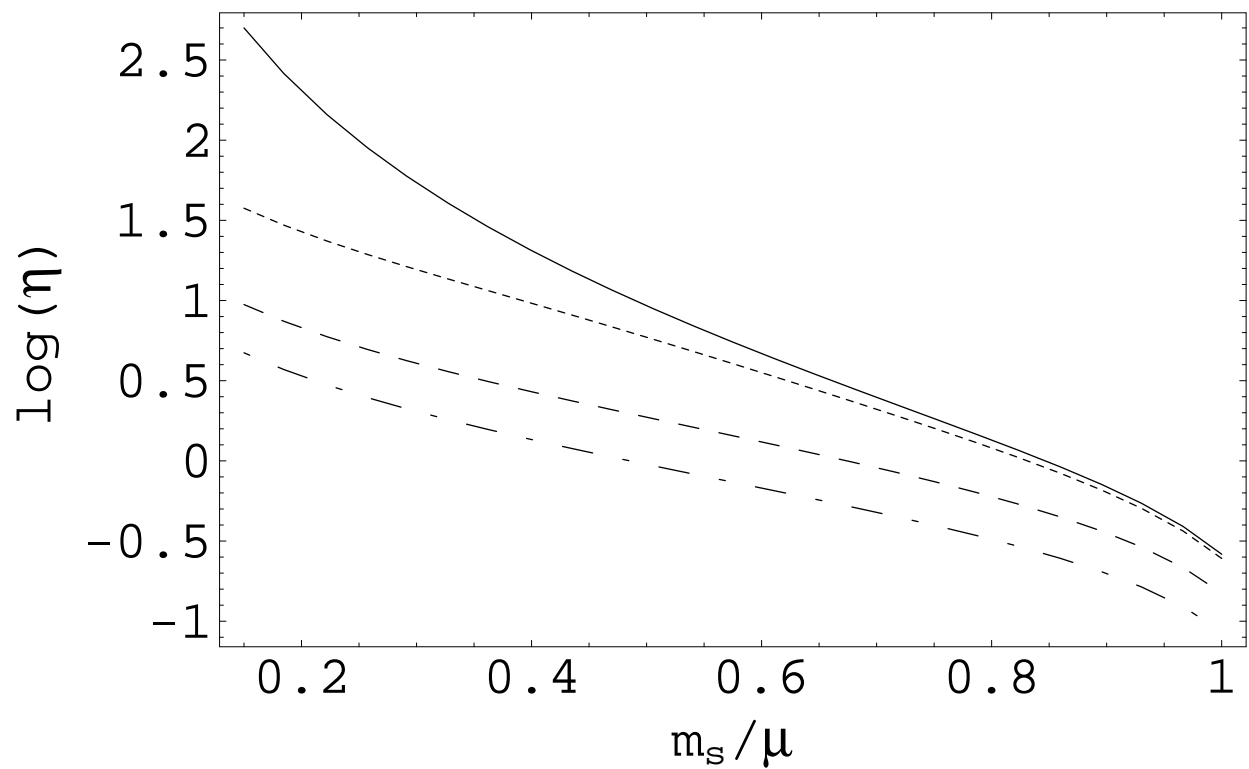} \caption{The parameter $\eta =\sigma /\sigma _0$
as a function of the ratio of the $s$ quark mass $m_s$ and the
average quark chemical potential $\mu$ for different temperatures
at the SQM surface: $T_S=0$ (solid curve), $T_S=5$ MeV (dotted
curve), $T_S=20$ MeV (dashed curve) and $T_S=40$ MeV
(dashed-dotted curve).} \label{FIG2}
\end{figure}

\vspace{0.2in}

\begin{figure}[h]
\plotone{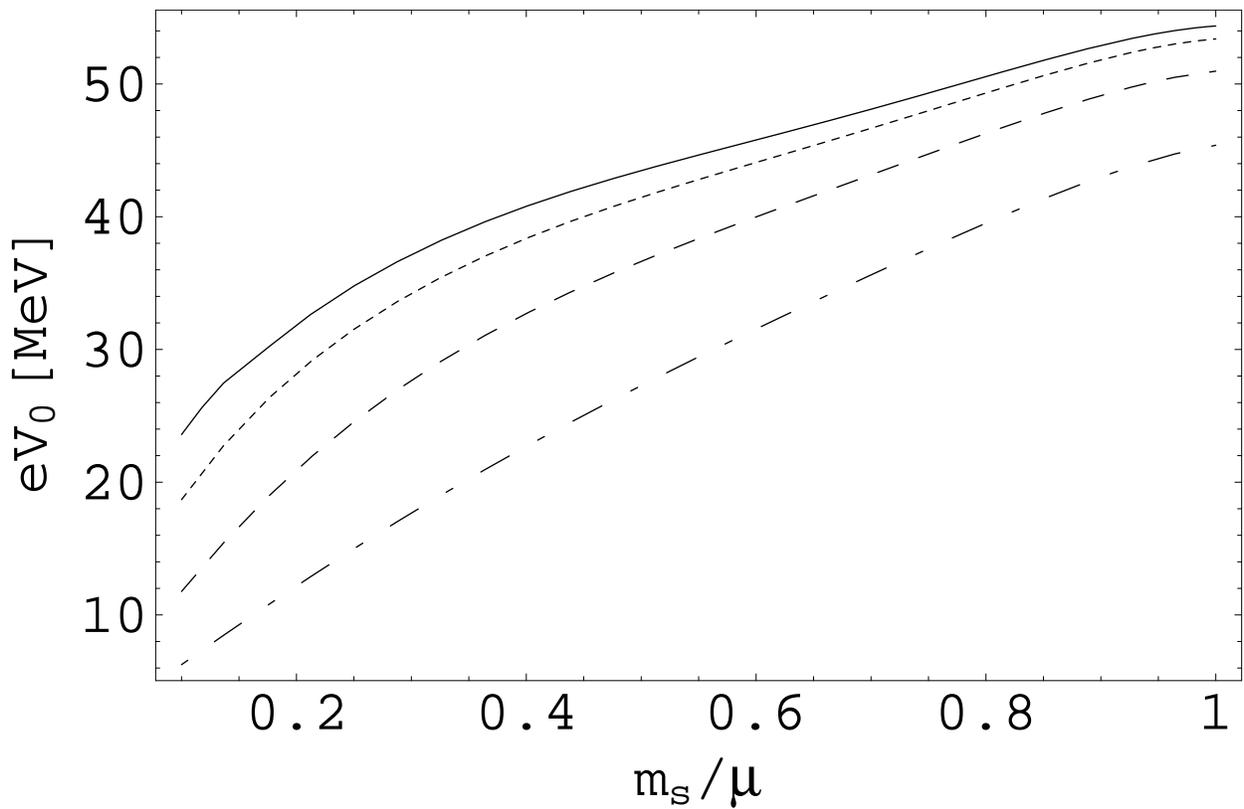} \caption{Electrostatic potential of electrons
$eV_0$ at the surface of unpaired SQM as a function of the ratio
of the $s$ quark mass $m_s$ and the average quark chemical
potential $\mu$ for $\mu=300$ MeV and different temperatures at
the SQM surface: $T_S=0$ (solid curve), $T_S=5$ MeV (dotted
curve), $T_S=10$ MeV (dashed curve) and $T_S=20$ MeV (
dashed-dotted curve).} \label{FIG3}
\end{figure}

\vspace{0.2in}

\begin{figure}[h]
\plotone{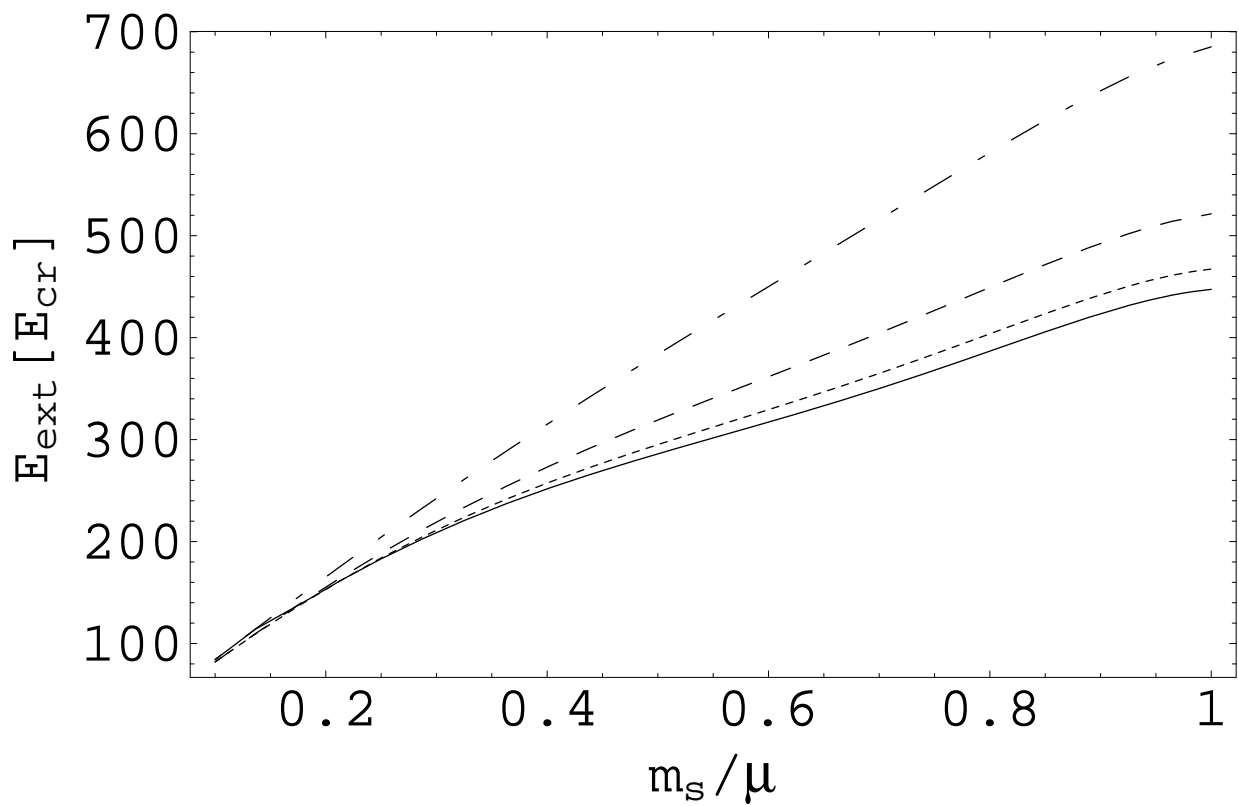} \caption{External electric field (in units of
$E_{\rm cr}$) at the SQM surface as a function of the ratio of the
mass of $s$ quarks and the average quark chemical potential $\mu$
for $\mu=300$ MeV and different temperatures at the SQM surface:
$T_S=0$ (solid curve), $T_S=5$ MeV (dotted curve), $T_S=10$ MeV
(dashed curve) and $T_S=20$ MeV ( dashed-dotted curve).}
\label{FIG4}
\end{figure}

\vspace{0.2in}

\begin{figure}[h]
\plotone{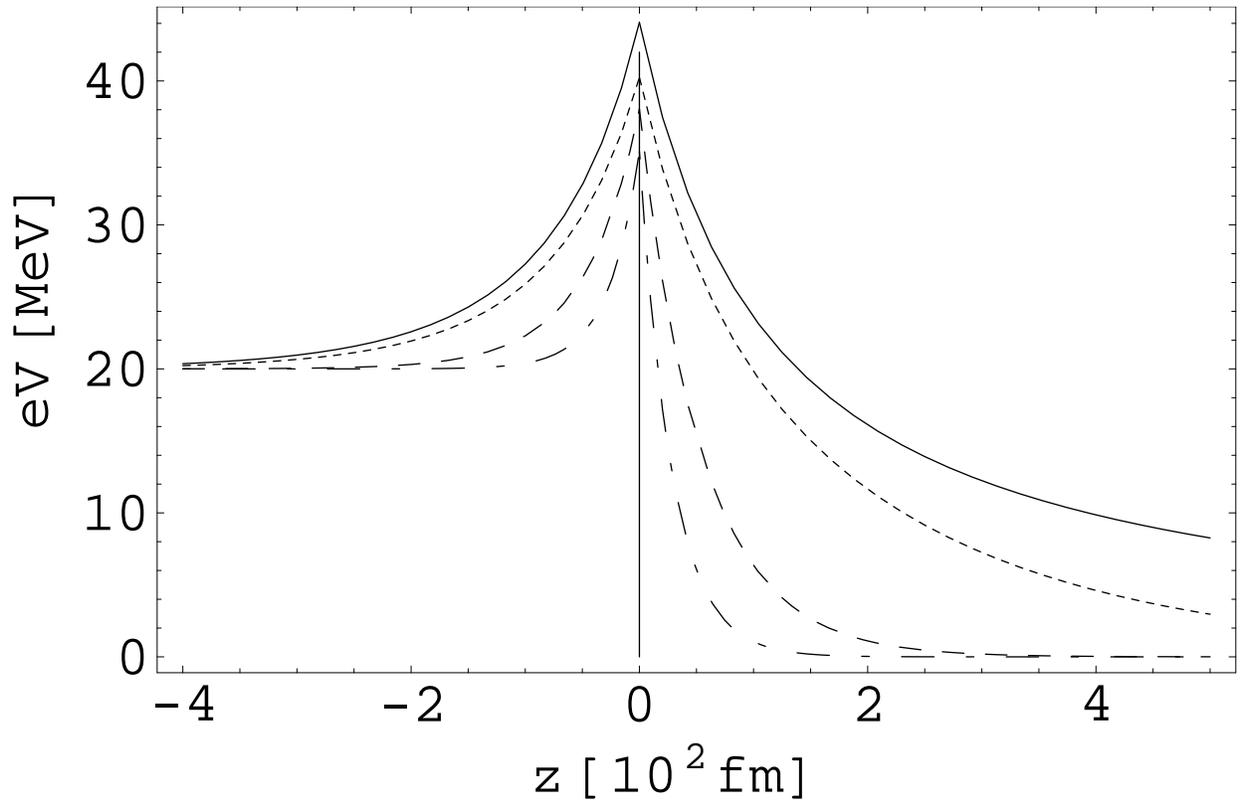} \caption{Electrostatic potential of electrons
$eV$ as a function of the distance $z$ from the SQM surface
($z=0$) for $m_s =150$ MeV, $\mu=300$ MeV, $eV_q=20$ MeV, and for
different temperatures at the surface: $T_S=0$ (solid curve),
$T_S=5$ MeV (dotted curve), $T_S=20$ MeV (dashed curve) and
$T_S=40$ MeV ( dashed-dotted curve).} \label{FIG5}
\end{figure}

\vspace{0.2in}

\begin{figure}[h]
\plotone{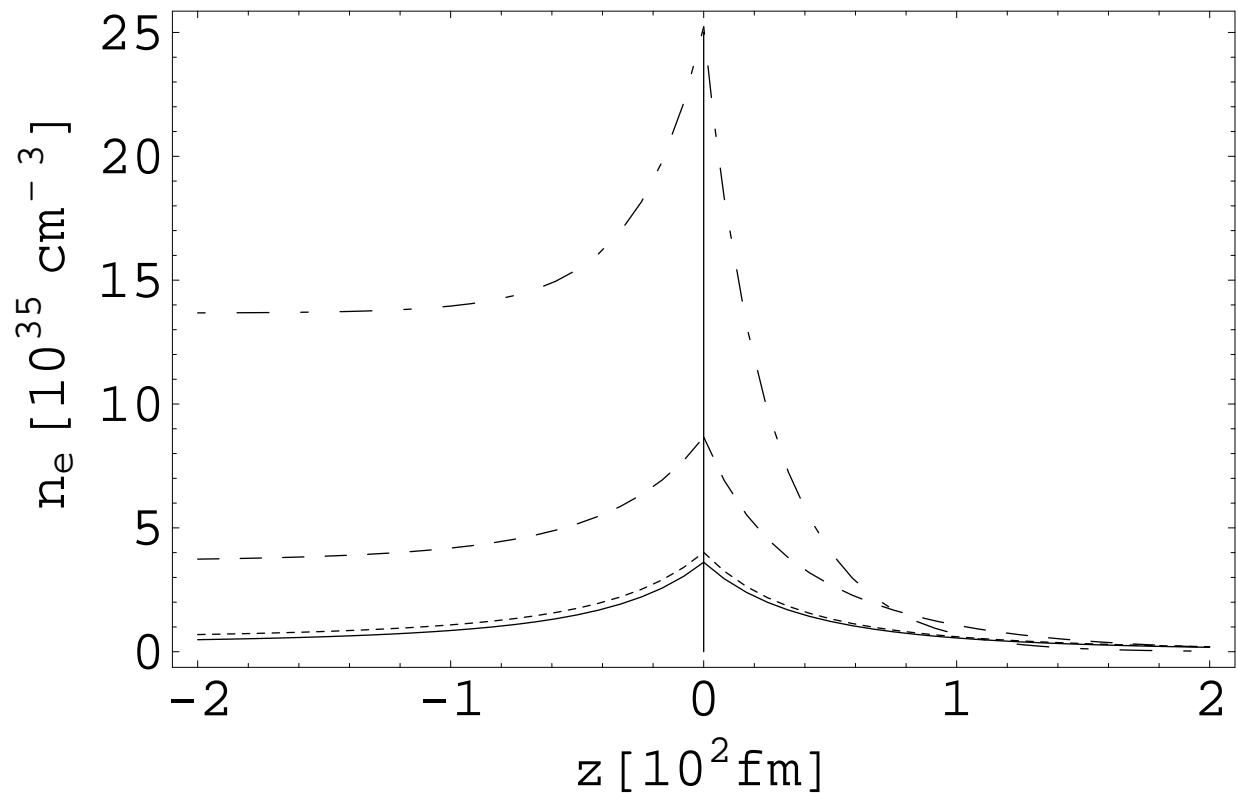} \caption{Electron number density profiles near
the SQM surface for $m_s =150$ MeV, $\mu=300$ MeV, $eV_q=20$ MeV,
and for different temperatures at the surface: $T_S=0$ (solid
curve), $T_S=5$ MeV (dotted curve), $T_S=20$ MeV (dashed curve)
and $T_S=40$ MeV ( dashed-dotted curve).} \label{FIG6}
\end{figure}


\newpage

\begin{thebibliography}{}
\bibitem[Alcock et al. 1986]{AFO86} Alcock, C., Farhi, E.,
\& Olinto, A. 1986, ApJ, 310, 261
\bibitem[Aksenov et al. 2003]{AMU03} Aksenov, A.G., Milgrom, M.,
\& Usov, V.V. 2003, MNRAS, 343, L69
\bibitem[Aksenov et al. 2004]{AMU04} Aksenov, A.G., Milgrom, M.,
\& Usov, V.V. 2004, ApJ, 609, 363
\bibitem[Alford 2001]{A01} Alford, M. 2001, Annu. Rev. Nucl. Part. Sci.,
51, 131
\bibitem[Alford et al. 2001a]{ABR01a} Alford, M., Bowers, J.A.,
\& Rajagopal, K. 2001a, J. Phys. G., 27, 541
\bibitem[Alford et al. 2001b]{ABR01b} Alford, M., Bowers, J.A.,
\& Rajagopal, K. 2001b, Phys. Rev. D, 63, 074016
\bibitem[Alford \& Reddy 2003]{AR03} Alford, M., \& Reddy, S. 2003,
Phys. Rev. D, 67, 074024
\bibitem[Alford et al. 2004]{AKR04} Alford, M., Kouvaris, C., \&
Rajagopal, K. 2004, Phys. Rev. Lett., 92, 222001
\bibitem[Alford \& Rajagopal 2002]{AR02} Alford, M., \& Rajagopal, K.
2002, J. High Energy Phys., 0206, 031
\bibitem[Alford et al. 1998]{ARW98} Alford, M., Rajagopal, K., \&
Wilczek, F. 1998, Phys. Lett. B, 422, 247
\bibitem[Bailin \& Love 1984]{BL84} Bailin, D., \& Love, A. 1984,
Phys. Rep., 107, 325
\bibitem[Baym et al. 1971]{BPS71} Baym, G., Pethick, C.J., \&
Sutherland, P.G. 1971, ApJ, 170, 299
\bibitem[Berger 1991]{B91} Berger, M.S. 1991, Phys. Rev. C,
44, 566
\bibitem[Berger \& Jaffe 1987]{BJ87} Berger, M.S., \& Jaffe, R.L.
1987, Phys. Rev. C, 35, 213
\bibitem[Bodmer 1971]{B71} Bodmer, D. 1971, Phys. Rev. D., 4, 1601
\bibitem[Bowers \& Rajagopal 2002]{BR02} Bowers, J.A., \&
Rajagopal, K. 2002, Phys. Rev. D, 66, 065002
\bibitem[Cheng et al. 1998]{CDL98} Cheng, K.S., Dai, Z.G., \& Lu, T.
1998, Int. J. Mod. Phys. D, 7, 139
\bibitem[Cheng \& Harko 2000]{CH00} Cheng, K.S., \& Harko, T. 2000,
Phys. Rev. D, 62, 083001
\bibitem[Cheng \& Harko 2003]{CH03} Cheng, K.S., \& Harko, T. 2003,
ApJ, 596, 451
\bibitem[Evans et al. 2000]{E00} Evans, N., Hormuzdiar, J.,
Hsu, S.D.H., \& Schwetz, M. 2000, Nucl. Phys. B, 581, 391
\bibitem[Farhi \& Jaffe 1984]{FJ84}
Farhi, E., \& Jaffe, R.L. 1984, Phys. Rev. D, 30, 2379
\bibitem[Glendenning 1996]{G96} Glendenning, N.K. 1996, Compact Stars:
Nuclear Physics, Particle Physics, and General Relativity (New
York: Springer)
\bibitem[Haensel et al. 1986]{HZS86} Haensel, P., Zdunik, J.L.,
\& Schaeffer, R.1986, ApJ, 160, 121
\bibitem[Hu \& Xu 2002]{HX02} Hu, J., \& Xu, R.X. 2002, A\&A,
387, 710
\bibitem[Huang \& Shovkovy 2003]{HS03} Huang, M., \& Shovkovy, I. 2003,
Nucl. Phys. A, 729, 835
\bibitem[Jaikumar et al. 2004]{JGPP04} Jaikumar, P., Gale, C.,
Page, D., \& Prakash, M. 2004, Phys. Rev. D, 70, 023004
\bibitem[Kettner et al. 1995]{KWWG95} Kettner, Ch., Weber, F.,
Weigel, M.K., \& Glendenning, N.K. 1995, Phys. Rev. D, 51, 1440
\bibitem[Lugones \& Horvath 2003]{LH03} Lugones, G., \& Horvath, J.E.
2003, A$\&A$, 403, 173
\bibitem[Madsen 2000]{M00} Madsen, J. 2000, Phys. Rev. Lett.,
85, 4687
\bibitem[Madsen 2001]{M01} Madsen, J. 2001, Phys. Rev. Lett.,
87, 172003
\bibitem[Page \& Usov 2002]{PU02} Page, D., \& Usov, V.V. 2002,
Pys. Rev. Lett., 89, 131101
\bibitem[Rajagopal \& Wilczek 2001]{RW01} Rajagopal, K., \&
Wilczek, F. 2001, Phys. Rev. Lett., 86, 3492
\bibitem[Rapp et al. 1998]{R98} Rapp, R., Sch$\ddot{\rm a}$fer, T.,
Shuryak, E.V., \& Velkovsky, M. 1998, Phys. Rev. Lett., 81, 53
\bibitem[Schafer 2000]{S00} Sch$\ddot{\rm a}$fer, T. 2000,
Nucl. Phys. B, 575, 269
\bibitem[Schwinger 1951]{S51} Schwinger, J. 1951, Phys. Rev., 82, 664
\bibitem[Shovkovy \& Huang 2003]{SH03} Shovkovy, I., \& Huang, M. 2003,
Phys. Lett. B, 564, 205
\bibitem[ Steiner et al. 2002]{SRP02} Steiner, A.W., Reddy, S.,
\& Prakash, M. 2002, Phys. Rev. D, 66, 094007
\bibitem[Usov 1998]{U98} Usov, V.V. 1998, Phys. Rev. Lett., 80, 230
\bibitem[Usov 2001a]{U01} Usov, V.V. 2001a, ApJ, 550, L179
\bibitem[Usov 2001b]{U01b} Usov, V.V. 2001b, Phys. Rev. Lett., 87, 021101
\bibitem[Usov 2004]{U04} Usov, V.V. 2004, Phys. Rev. D, 70,
067301
\bibitem[Weber 1999]{W99} Weber, F. 1999, J. Phys. G, 25, 195
\bibitem[Weber 2004]{W04} Weber, F. 2004, astro-ph/0407155
\bibitem[Witten 1984]{W84} Witten, E. 1984, Phys. Rev. D, 30, 272

\end{thebibliography}
\end{document}